\documentclass[aps,prl,superscriptaddress,nofootinbib]{revtex4}

\usepackage[centertags,sumlimits,intlimits,namelimits,reqno]{amsmath}
\usepackage{bbm,amsfonts}
\usepackage{amsthm}

\usepackage{graphicx}
\usepackage{dcolumn}
\usepackage{bm}

\newcommand{\ud}[2]{\genfrac{}{}{0pt}{1}{#1}{#2}}

\newcolumntype{S}{>{\scancell}l<{\endscan}}
\def\scancell#1\ignorespaces#2\unskip\endscan{%
  $\mathsf{#2}$%
}

\theoremstyle{plain}
\newtheorem{theorem}{Theorem}

\newtheorem{fact}[theorem]{Fact}

\DeclareMathOperator{\tr}{tr}

\def\Tr{\mathrm{Tr}}
\def\tr{\mathrm{tr}}

\def\E{{\mathcal E}}

\def\>{\rangle}
\def\<{\langle}

\def\bfact{\begin{fact}}
\def\efact{\end{fact}}

\def\bv{\left( \begin{matrix}}
\def\ev{\end{matrix} \right)}

\def\be{\begin{equation}}
\def\ee{\end{equation}}
\def\bes{\begin{eqnarray}}
\def\ees{\end{eqnarray}}
\def\bess{\begin{eqnarray*}}
\def\eess{\end{eqnarray*}}

\newcommand{\Cov}{{\mathrm{Cov}}}
\newcommand{\e}{{\mathrm e}}

\begin{document}

\title{
Symmetrised Characterisation of Noisy Quantum Processes}

\author{Joseph Emerson}
\affiliation{Institute for Quantum Computing, University of
Waterloo} \affiliation{Department of Applied Mathematics, University
of Waterloo}
\author{Marcus Silva}\affiliation{Institute
for Quantum Computing, University of
Waterloo}\affiliation{Department of Physics and Astronomy,
University of Waterloo}
\author{Osama Moussa}\affiliation{Institute
for Quantum Computing, University of Waterloo}
\affiliation{Department of Physics and Astronomy, University of
Waterloo}
\author{Colm Ryan}\affiliation{Institute for Quantum Computing, University of
Waterloo}\affiliation{Department of Physics and Astronomy,
University of Waterloo}
\author{Martin Laforest}
\affiliation{Institute for Quantum Computing, University of
Waterloo} \affiliation{Department of Physics and Astronomy,
University of Waterloo}
\author{Jonathan Baugh}\affiliation{Institute for Quantum Computing, University of
Waterloo}\affiliation{Department of Physics and Astronomy,
University of Waterloo}
\author{David G. Cory}\affiliation{Department of Nuclear Science and Engineering, Massachusetts Institute of Technology}
\author{Raymond Laflamme}\affiliation{Institute for Quantum Computing, University of
Waterloo}\affiliation{Department of Physics and Astronomy,
University of Waterloo}

\date{\today}

\begin{abstract}
A major goal of developing high-precision control of many-body
quantum systems is to realise their potential as quantum computers.
Probably the most significant obstacle in this direction is the
problem of "decoherence": the extreme fragility of quantum systems
to environmental noise and other control limitations. The theory of
fault-tolerant quantum error correction has shown that quantum
computation is possible even in the presence of decoherence provided
that the noise affecting the quantum system satisfies certain
well-defined theoretical conditions. However, existing methods for
noise characterisation have become intractable already for the
systems that are controlled in today's labs. In this Report we
introduce a technique based on symmetrisation that enables direct
experimental characterisation of key properties of the decoherence
affecting a multi-body quantum system. Our method reduces the number
of experiments required by existing methods from exponential to
polynomial in the number of subsystems. We demonstrate the
application of this technique to the optimisation of control over
nuclear spins in the solid state.
\end{abstract}

\maketitle

Quantum information enables efficient solutions to certain tasks
which have no known efficient solution in the classical world. This
discovery has reshaped our understanding of computational complexity
and emphasized the physical nature of information. A necessary
condition to take advantage of the quantum world is the ability to
gain robust control of quantum systems and, in particular,
counteract the noise and decoherence affecting any physical
realisation of quantum information processors (QIPs). A pivotal step
in this direction came with the discovery of quantum error
correction  codes (QECCs)\cite{Sho95a,Ste96a} and the associated
accuracy threshold theorem for fault-tolerant (FT) quantum
computation
\cite{shor:qc1996a,aharonov:qc1996a,kitaev:qc1997a,knill:qc1998a}.
In order to make use of quantum error correction and produce
fault-tolerant protocols, we need to understand the nature of the
noise affecting the system at hand. There is a direct way to fully
characterise the noise using a procedure known as process tomography
\cite{NC-QPT,AAPT,DC}. However, this procedure requires resources
that grow exponentially with the number of subsystems (usually
two-level systems called `qubits'). As a result, process tomography
is an intractable procedure for characterizing the multi-qubit
quantum systems that have already been realised
\cite{8qubit,Mqubits,12qubit}. In this letter we introduce a general
symmetrisation method that allows for direct experimental
characterisation of relevant features of the noise. We apply this
framework to develop an efficient experimental protocol for
characterising multi-qubit correlations and memory effects in the
noise. Compared to existing methods \cite{NC-QPT,AAPT,DC}, the
protocol yields an exponential savings in the number of experiments
required to obtain such information. In the context of applications,
this information enables tests of some key assumptions underlying
estimates of the FT threshold and optimisation of error-correction
strategy. Moreover, the estimated noise parameters are immediately
relevant for optimizing experimental control methods. We demonstrate
this optimization through an implementation of the protocol on a
solid-state nuclear magnetic resonance (NMR) QIP.

\begin{figure}[ht]
\includegraphics{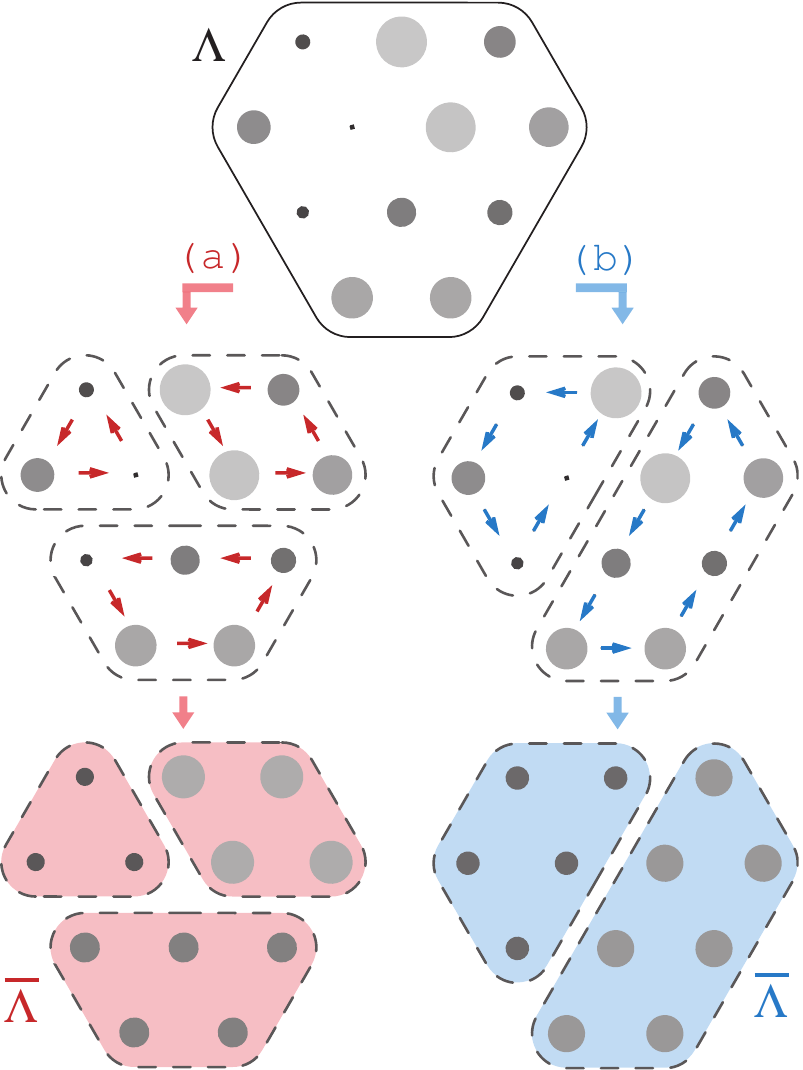}
\caption{\label{figAlphas} \textbf{Schematic illustration of
coarse-graining via symmetrization.} A general quantum process
$\Lambda$ is described by a finite but very large number of
independent parameters (represented by shaded circles). Averaging
the map by conjugation under sets of operators drawn from an
appropriate symmetry group yields an effective quantum process which
has a reduced number of independent parameters.  In the figure we
illustrate the results of two different symmetrization groups
(represented by (a) red and (b) blue) which take the parameters
around closed orbits. The reduced (coarse-grained) parameters of the
averaged quantum process $\overline{\Lambda}$ can be determined by
selecting a set of initial states $\rho_i$ such that the output
states $\overline{\Lambda}(\rho_i)$ carry a signature of the reduced
parameters.  In this work, we design and implement a symmetrization
protocol that is relevant for characterizing the performance of
quantum error correcting codes and testing some of the assumptions
of fault-tolerance threshold theorems.}
\end{figure}
\begin{figure}[ht]
\includegraphics{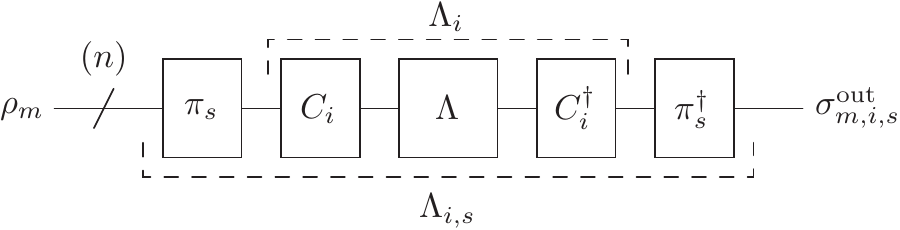}
\caption{\label{figCirc} \textbf{Schematic quantum circuit} for one
experimental run consisting of a conjugation of the noise process
$\Lambda$. Time flows from left to right. The standard protocol
requires conjugation only by an element $C_i \in
\mathcal{C}_1^{\otimes n}$, whereas the ensemble protocol requires
conjugating $\Lambda$ also by a permutation $\pi_s \in
\mathcal{S}_{n,w'} \subset \mathcal{S}_n$ of the $n$ qubits. The
standard protocol requires only one (pure) input state $| 0
\rangle^{\otimes n}$, whereas the ensemble protocol requires $n$
distinct input operators $\rho_{w'}= Z^{\otimes w'} \otimes
\openone^{\otimes (n-w')}$ with $w' \in \{1,...,n\}$.  In either
case the protocol involves running this circuit $k =
\mathcal{O}(\log(2(n+1))/\delta^2)$ times for each input
configuration to estimate the $n+1$ output parameters to precision
$\delta$ with constant probability.}
\end{figure}

 We focus here on the noise affecting a system of $n$ qubits.
 The crucial point is that a
complete description of a general noise model $\Lambda$ requires
$O(2^{4n})$ parameters. Clearly an appropriate coarse-graining of
this information is required; the challenge is to identify
\emph{efficient} methods for estimating the features of practical
interest. The method we propose is based on identifying a symmetry
associated with the properties of interest, and then operationally
symmetrizing the noise process to yield an effective map
$\overline{\Lambda}$ with a reduced number of independent parameters
that reflect these properties (see Fig.~\ref{figAlphas}). This
symmetrization is achieved by conjugating the map with a unitary
operator drawn from the relevant symmetry group (see
Fig.~\ref{figCirc}) and then averaging with respect to that group
\cite{Emerson03,LopezLevi06,EAZ05,DCEL,Cirac05}. As described below,
rigorous bounds guarantee that the number of experimental trials
required remains independent of the dimension of the group. Hence
the randomization method leads to efficient partial characterization
of the map $\Lambda$ whenever the group elements admit efficient
circuit decompositions.

We apply this general idea to the important problem of estimating
the noise parameters that determine the performance of a broad class
of QECCs and the applicability of certain assumptions underlying FT
thresholds. In general, QECCs protect quantum information only
against certain types of noise. A distance-$(2t+1)$ code refers to
class of codes that correct all errors simultaneously affecting up
to $t$ qubits. Hence the distance of an error correcting code
determines which terms in the noise process will be corrected and
which will remain uncorrected. The latter contribute to the overall
failure probability. Of course, it is possible to estimate the
failure probability under the \emph{assumption} that the noise is
independent from qubit to qubit \cite{knill:qc1998a} or between
blocks of qubits. Many fault-tolerance theorems assume this kind of
behavior. Hence a fundamental problem is to measure the correlations
in the noise for a given experimental arrangement \emph{without} the
exponential overhead of process tomography. We report here a
protocol that achieves this goal. We also show that this protocol
remains efficient also in the context of an ensemble QIP with highly
mixed states \cite{NMRQIP}.

We start by expanding the noise operators in a basis of operators
$P_i \in \mathcal{P}_n$, which consist of $n$-fold tensor product of
the usual single-qubit Pauli operators $\{ 1,X,Y,Z \}$ satisfying
the orthogonality relation $\Tr[P_i P_j] = D \delta_{ij}$. The
Clifford group $\mathcal{C}_n$ is defined as the normalizer of the
Pauli group $\mathcal{P}_n$: it consists of all elements $U_i$ of
the unitary group $U(D)$ satisfying $ U_i P_j U_i^\dagger \in
\mathcal{P}_n $ for every $P_j \in \mathcal{P}_n$. The protocol
requires symmetrizing the channel $\Lambda \rightarrow
\overline{\Lambda} $ by averaging over trials in which the channel
is conjugated by the elements of $\mathcal{C}_1$ applied
independently to each qubit (see Fig.~\ref{figCirc}). An average
over conjugations is known as a ``twirl'' \cite{BDSW,DCEL}, and
hence the above is a $\mathcal{C}_1^{\otimes n}$-twirl.

Separating out terms according to their Pauli weight $w$, where $w
\in \{0,\dots, n\} $ is the number of non-identity factors in $P_l$,
letting the index $\nu_w \in \{1, \dots, {n \choose w} \}$ count the
number of distinct ways that $w$ non-identity Pauli operators can be
distributed over the $n$ factor spaces and the index $\mathbf{i}_w =
\{i_1, \dots, i_w \}$ with $ i_j \in \{1, 2, 3 \}$ denote which of
the non-identity Pauli operators occupies the $j$'th occupied site,
we obtain
\be \overline{\Lambda(\rho)} = \sum_{w=0}^n
\sum_{\nu_w=1}^{{n \choose w}}
    r_{w,\nu_w} \sum_{\mathbf{i}_w}
    P_{w,\nu_w,\mathbf{i}_w} \rho P_{w,\nu_w,\mathbf{i}_w}
\ee
where $ r_{w,\nu_w} = \frac{1}{3^w}  \sum_{\mathbf{i}_w}
a_{w,\nu_w,\mathbf{i}_w}$. (Details of this derivation are given in
the appendix.) If the symmetrized channel is probed by the initial
state $|0\> \equiv | 0\>^{\otimes n}$, followed by a projective
measurement of the output state in the basis $|l\>$, this yields an
$n$-bit string $l \in \{ 0, 1 \}^n $. Let $q_{w'}$ denote the
probability that a random subset of $w'$ bits of the binary string
$l$ has even parity. Noting that $c_{w'} \equiv \langle
\overline{Z^{\otimes w'}} \rangle = 2q_{w'} - 1$, where
$\overline{Z^{\otimes w'}}$ is the average of all Pauli operators
with $w'$ factors of $Z$ and $n-w'$ identity factors, we obtain $p_w
= \sum_{w'} \Omega_{w,w'}^{-1} c_{w'}$ where the matrix
$\Omega_{w,w'}^{-1}$ is a matrix of combinatorial factors given in
the appendix, and the $p_w$ are the probabilities of $w$
simultaneous qubit errors occurring over the course of the quantum
process. Of course all imperfections in the protocol contribute to
the total probabilities of error. The protocol can be made robust
against imperfections in the input state preparation, measurement
and twirling by substituting $ c_{w'} \rightarrow \tilde{c}_{w'} =
c_{w'}/\langle \overline{Z^{\otimes w'}}\rangle_{0}$, where we
simply factor out the expectation value observed when the protocol
is performed without the noisy channel. In this case the
probabilities of different error weights are given by $ p_w =
\sum_{w'} \Omega_{w,w'}^{-1} \tilde{c}_{w'}.$

If in each single shot experiment the Clifford operators are chosen
uniformly at random then with $K =
\mathcal{O}(\log(2(n+1))/\delta^2)$ experiments we can estimate each
of the coefficients $c_w$ to precision $\delta$ with constant
probability. The $c_{w'}$ can be applied directly to test some of
the assumptions that yield rigorous estimates of the fault-tolerance
threshold \cite{AGP}. In particular, a noisy channel with an
uncorrelated distribution of error locations, but with arbitrary
correlations in the error type at each location, is mapped under our
symmetrization to a channel which is a tensor product of $n$
single-qubit depolarizing channels. A channel satisfying this
property will exhibit the scaling $c_w=c_1^w$. Hence observed
deviations from this scaling law imply a violation of the above
assumption. Furthermore, the question of whether the noise exhibits
non-Markovian properties  can be tested efficiently by repeating the
above scheme for distinct time-intervals $m\tau$ with increasing
$m$. Markovian noise is guaranteed to satisfy the semi-group
property $ \overline{\Lambda}_{\tau_1} \circ
\overline{\Lambda}_{\tau_2} = \overline{\Lambda}_{\tau_1+\tau_2},$
and hence memory effects in the noise are implied by any observed
deviations from the condition $c_w(m\tau) = (c_w(\tau))^m$.

The estimates $c_{w'}$ also give estimates for $p_w$, but the
statistical uncertainty for $p_w$ grows exponentially with $w$.
Using a bound on $\Omega_{w,w'}^{-1}$ derived in the appendix, we
show that all $p_w$ for which $w \leq l$ can be estimated with
$\mathcal{O}(n^l)$ trials. This allows for characterisation of other
important features of the noise. The probability $p_0$ is directly
related to the entanglement fidelity of the channel and hence this
protocol provides an exponential savings over recently proposed
methods for estimating this single figure of merit
\cite{EAZ05,Fortunato02,NielsenFidelity} (see also
Ref.~\cite{DCEL}). Hence, by actually implementing any given code we
can bound the failure probability of that code with only
$\mathcal{O}(\log(2n)/\delta^2)$ experiments without making any
theoretical assumptions about the noise. Moreover, on physical
grounds we may expect the noise to become independent between qubits
outside some fixed (but unknown) scale $b$, after which the $p_w$
decrease exponentially with $w$. The scale $b$ can be determined
efficiently with $n^b$ experiments.

While a characterisation of the twirled channel is useful given the
relevance of twirled channels to fault-tolerant applications
\cite{knill-nature,knill-PRA}, we remark that the failure
probability of the twirled channel gives an upper bound to the
failure probability of the original un-twirled channel whenever the
performance of the code has some bound that is invariant under the
symmetry associated with the twirl. This holds quite generally in
the context of the symmetry considered above because the failure
probability of a generic distance-$2t+1$ code is bounded above by
the total probability of error terms with Pauli weight greater than
$t$ and this weight remains invariant under conjugation by any
element $C_i \in \mathcal{C}_1^{\otimes n}$.

Our method provides an efficient protocol for the characterisation
of the noise in contexts where the target transformation is the
identity operator, e.g. a quantum communication channel or quantum
memory. However, the protocol also provides an efficient means for
characterising the noise under the action of a non-identity unitary
transformation such as a quantum gate. There are two ways to adapt
the protocol to this setting. First, we recall that a unitary
transformation can be decomposed into a product of basic quantum
gates drawn from a universal gate set, where each gate in the set
acts on at most $2$ qubits simultaneously. Hence, the noise map
acting on all $n$ qubits associated with any $2$-qubit gate can be
determined by applying the above protocol to other $n-2$ qubits
while applying process tomography to the $2$ qubits in the domain of
the quantum gate. A second approach is to estimate the average
error-per-gate for a sequence of $m$ gates such that the composition
gives the identity operator. Such a sequence can be generated by
making use of the cyclic property $U^m = 1$ of any gate in a
universal gate set, or by choosing a sequence of $m-1$ random gates
followed by an $m$'th gate chosen such that the composition gives
the identity transformation \cite{Knill}.

\renewcommand{\arraystretch}{1.5}
\begin{table}[ht!]
\begin{center}
\begin{tabular}{|c|l|l|c|c|c|c|c|c|}
\hline
\#
&
System
&
Map Description
&
Kraus operators ($A_k$)
&
$k$ & $p_0$ & $p_1$ & $p_2$ & $p_3$ \\
\hline
1
&
CHCl$_3$
&
Engineered: $\mathbf{p}=[0,1,0]$.
&
$\frac{1}{\sqrt2} \{Z_1, Z_2\}$
&
288 & 0.000 $\ud{+0.004}{}$ & 0.991 $\ud{+0.009}{-0.015}$ & 0.009 $\ud{+0.017}{-0.009}$ & - \\
2
&
CHCl$_3$
&
Engineered: $\mathbf{p}=[0,0,1]$.
&
$\{ Z_1 Z_2\}$
&
288 & 0.001 $\ud{+0.006}{-0.001}$ & 0.004 $\ud{+0.011}{-0.004}$ & 0.996 $\ud{+0.004}{-0.011}$ & - \\
3
&
CHCl$_3$
&
Engineered: $\mathbf{p}=[\frac{1}{4},\frac{1}{2},\frac{1}{4}]$.
&
$\{\exp[i \frac{\pi}{4} (Z_1+Z_2)] \}$
&
288 & 0.254 $\ud{+0.010}{-0.010}$ & 0.495 $\ud{+0.021}{-0.020}$ & 0.250 $\ud{+0.019}{-0.019}$ & - \\
\hline
4
&
C$_3$H$_4$O$_4$
&
Engineered: $\mathbf{p}=[0,1,0,0]$.
&
$\frac{1}{\sqrt3}\{ Z_1,Z_2, Z_3\}$
&
432 & 0.01 $\ud{+0.01}{-0.01}$ & 0.99 $\ud{+0.01}{-0.03}$ & 0.01 $\ud{+0.02}{-0.01}$ &0.00 $\ud{+0.01}{}$ \\
5
&
C$_3$H$_4$O$_4$
&
Natural noise (a)
&
unknown
&
432 & 0.44 $\ud{+0.01}{-0.02}$ & 0.45 $\ud{+0.03}{-0.03}$ & 0.10 $\ud{+0.04}{-0.08}$ & 0.01 $\ud{+0.03}{-0.01}$ \\
6
&
C$_3$H$_4$O$_4$
&
Natural noise (b)
&
unknown
&
432 & 0.84 $\ud{+0.01}{-0.01}$ & 0.15 $\ud{+0.02}{-0.03}$ & 0.01 $\ud{+0.03}{-0.01}$ & 0.00 $\ud{+0.02}{}$ \\
\hline
\end{tabular}
\caption{\label{tabresults} \textbf{Summary of experimental
results.} The first four sets of experiments (three sets on the
two-qubit liquid-state system, and one on the three-qubit
solid-state system) were designed to characterise the performance of
the protocol under engineered noise. The last two sets were designed
to characterize the (unknown) natural noise due to imperfect
averaging of the internal Hamiltonian terms under a multiple-pulse
time-suspension sequence \cite{C48} with (a) one cycle with 10$\mu
s$ pulse-spacing, and (b) two cycles with 5$\mu s$ pulse-spacing.
The latter experiments characterized the residual errors under the
sequence with respect to the particular symmetry implemented by the
twirling process. The results of the solid-state experiments are
calculated after factoring out the effect of decoherence occurring
during the twirling operations (which only accounted for $\simeq
1\%$ fidelity loss per gate) in order to isolate the noise
associated with the time-suspension sequence.}
\end{center}
\end{table}%

\begin{figure}[ht]
\includegraphics{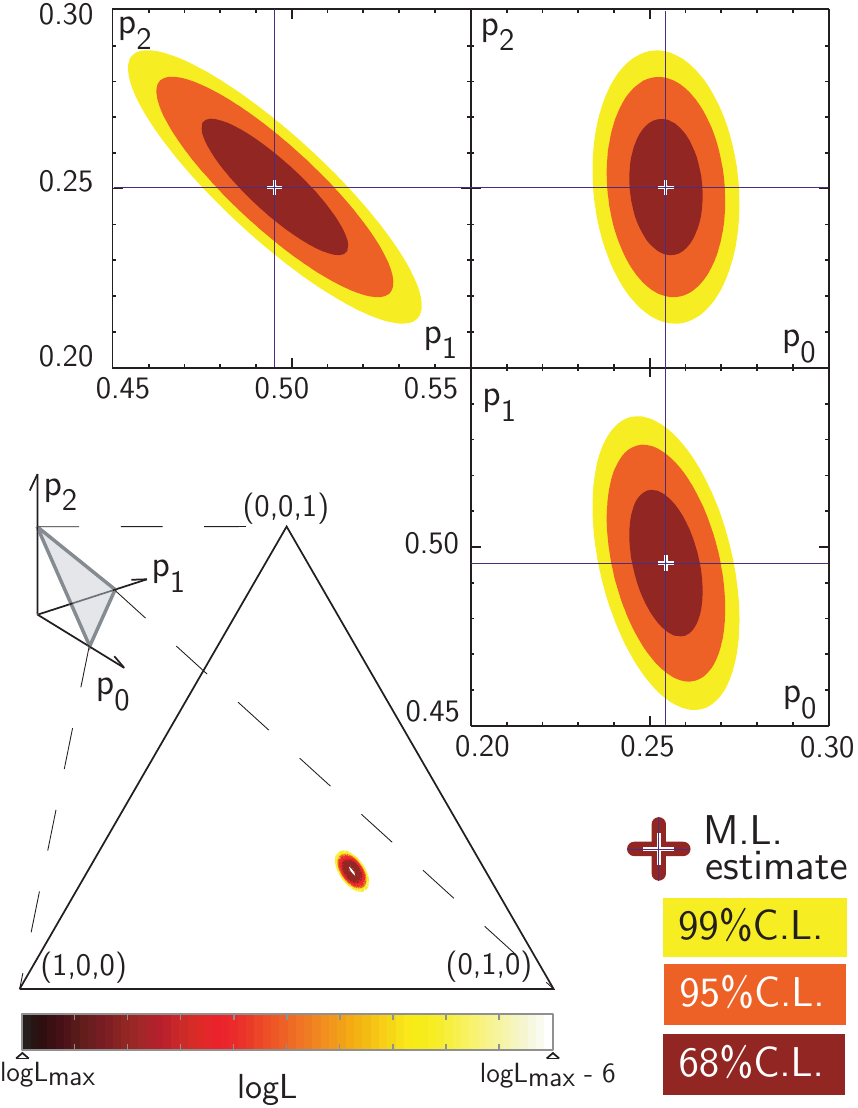}
\caption{\label{figResults} \textbf{Results for experiment \#3 in
Table \ref{tabresults}.} Shown is the Maximum Likelihood (ML)
estimate, $\mathbf{p}_{exp}=[ 0.254 \ud{+0.010}{-0.010} , 0.495
\ud{+0.021}{-0.020}, 0.250 \ud{+0.019}{-0.019}]$, for the error
probabilities of the engineered noise, $A_1=\{\exp[-i \frac{\pi}{4}
(Z_1+Z_2)] \}$, for which the calculated values are
$\mathbf{p}=[\frac{1}{4},\frac{1}{2},\frac{1}{4}]$. Also shown are
the confidence regions for the 68\%, 95\%, and 99\% confidence
levels (C.L.), which can be determined from the log of a likelihood
function, $\log L$. The experiment was performed on a 2-qubit
liquid-state NMR processor, and the noise was implemented by
appropriately phase-shifting the pulses. These experiments
illustrate the precision with which the protocol can be implemented
under conditions of well-developed quantum control.}
\end{figure}

We now describe how the above protocol is efficient also in the
context of an ensemble QIP \cite{NMRQIP}. First, we prepare
deviations from the identity state of the form $\rho_{w'} =
Z^{\otimes w'} \otimes \openone^{\otimes (n-w')}$ with $w'=\{1,
...\; n\}$. Hence the (non-scalable) preparation of pseudo-pure
states is not required. Second, we directly measure $\langle
\overline{Z^{\otimes w'}} \rangle$ for each $w'$ from the
expectation value $\langle Z^{\otimes w'} \openone^{\otimes n-w'}
\rangle$ by explicitly performing a random permutation of the
qubits. Let $\mathcal{S}_{n,w'}$ denote any subset of the group of
permutations of $n$ qubits, $\mathcal{S}_n$, such that $ \sum_{s}
\pi_{s}\rho_{w'}\pi_{s}^{\dagger} = \sum_{\nu_{w'}}
\rho_{w',\nu_{w'}} $, where $\nu_{w'} \in \{1, ... , \, {n \choose
w'} \}$ denotes each distribution of the $w'$ $Z$-Pauli operators.
As illustrated in Fig.~\ref{figCirc}, the symmetrisation consists of
conjugating the process $\Lambda$ with  $C_i\in
\mathcal{C}_1^{\otimes n}$ and $\pi_s \in \mathcal{S}_{n,w'}$. Let
$\Lambda_{i,s}$ stand for the conjugated noise-map, then, given
input operator $\rho_{w'}$, the output is
$\sigma^{\textrm{out}}_{w',i,s} = \Lambda_{i,s}(\rho_{w'})$.
Averaging the output operators $\sigma^{\textrm{out}}_{w',i,s}$ over
$i$ and $s$  gives the input operator scaled by $c_{w'}$.

\begin{figure}[ht]
\includegraphics{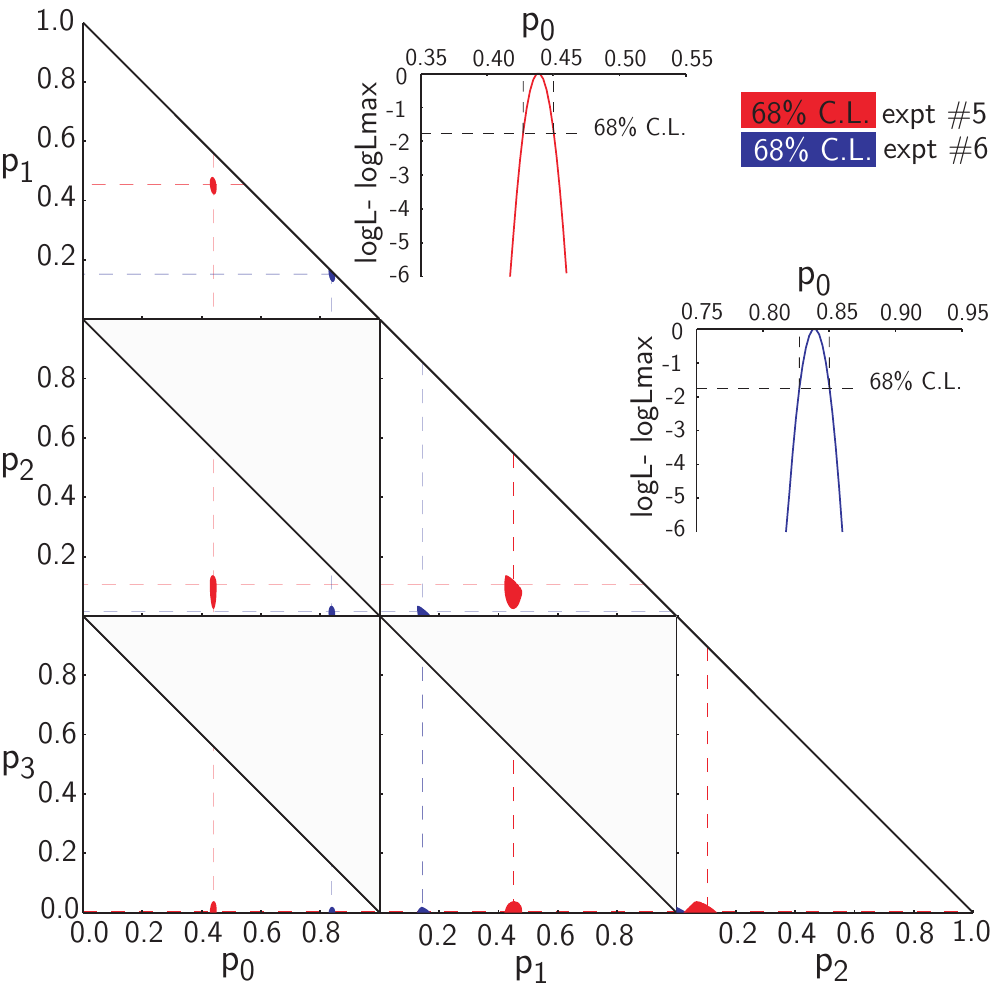}
\caption{\label{figSSResults} \textbf{Results for experiments \#5
and \#6 in Table \ref{tabresults}.} Shown are projections of the
4-dimensional likelihood function onto various probability planes.
The asymmetry seen in some of the confidence areas is a result of
this projection. The results for one cycle with 10$\mu$s
pulse-spacing (exp \#5) are in red and the results for  two cycles
with 5$\mu$s spacing (exp \#6) are in blue. The rate set by a
pulse-spacing of 10$\mu$s leads to a poor overall fidelity for the
time-suspension sequence, and, in particular, a significant
probability of two-qubit errors from the residual terms in the
effective Hamiltonian. Under two repetitions of the sequence with
the pulse-spacing of 5$\mu$s (which has twice as many pulses as the
single sequence with the 10$\mu$s spacing), the probability of
two-qubit errors in the residual terms decreases substantially. The
overall fidelity of either pulse-sequence is still limited by
incomplete heteronuclear decoupling of the qubits (carbon nuclei)
from the environment (nearby hydrogen nuclei): at the 5$\mu$s
pulse-spacing the carbon nuclei are being modulated on a time-scale
close to the proton decoupling frequency, so that the decoupling
sequence no longer averages the heteronuclear coupling to zero. A
simple extension of the protocol allows for determining which qubit
accrues the greatest single qubit errors from this effect. The
results were consistent with identifying the primary source of
single-qubit errors as the qubit whose coupling to the hydrogen is
an order of magnitude larger than that of the other two. By
increasing the decoupling power we were able to reduce these errors
though they could not be removed completely due to hardware
limitations.}
\end{figure}

We illustrate an implementation of the above protocol on both a
2-qubit (chloroform CHCl$_3$) liquid-state and a 3-qubit
(single-crystal Malonic acid C$_3$H$_4$O$_4$) solid-state NMR QIP
(see the appendix for more information on methods).
 The results of six sets of experiments are summarized in Table \ref{tabresults},
and detailed results for one liquid-state set are shown in
Fig.~\ref{figResults} and for two solid-state sets are shown in
Fig.~\ref{figSSResults}. The first four sets of experiments were
performed under engineered noise to both characterize the
performance of the protocol and to confirm high-fidelity control.
Two final two sets of experiments were performed in the solid-state
to characterize the unknown residual noise occurring under (a) one
cycle of a C48 pulse sequence \cite{C48} with 10$\mu s$ pulse
spacing, and (b) two cycles of C48 with 5$\mu s$ pulse spacing. The
C48 sequence is designed to suppress the dynamics due to the
system's internal Hamiltonian. The evolution of the system under
this pulse sequence can be evaluated theoretically by calculating
the Magnus expansion \cite{WaHa} of the associated effective
Hamiltonian, under which the residual effects appear as a sum of
terms associated with the Zeeman and dipolar parts of the
Hamiltonian, including cross terms. Roughly speaking, effective
suppression of the $k^{\mathrm{th}}$ term of the Hamiltonian takes
places when $\gamma_k \tau_k \ll 1$, where $\gamma_k$ is the
strength of the term and $\tau^{-1}_k$ is the rate at which it is
modulated by the pulse sequence. Generally shorter delays lead to
improved performance unless there is a competing process at the
shorter time-scale or there are limitations due to pulse
imperfections. Hence the performance of the pulse sequence is best
evaluated via experimental characterisation. The results shown in
Table \ref{tabresults} and Fig. \ref{figSSResults} illustrate how
the protocol can compare the performance of the two multiple-pulse
time-suspension sequences both in terms of the overall fidelity and
in terms of the relative probability of one, two and three body
noise terms.

The method above is an illustration of the first step in a hierarchy
of tests that are available under the general approach, with each
test giving more fine-grained information. For example, a variation
of the protocol in which only a Pauli-twirl is applied enables a
estimation of the $\mathcal{O}(n^3)$ relative probabilities of Pauli
$X$, $Y$ and $Z$ errors. This more fine-grained information is
useful not only for optimisation over QECC in eventual applications,
but also to characterize current performance of a given experimental
set-up. The full scope of information that can be estimated
efficiently via this general symmetrisation approach is an important
topic for further research. A specific question in this direction is
to determine whether this approach might enable efficient detection
of the presence of noiseless subsystems.

\appendix

\section{Appendix A: Analysis of the Symmetrisation}

The generic noise affecting a quantum state $\rho$ (a positive
matrix of dimension $D \times D$) can be represented by a completely
positive map of the form $\Lambda(\rho) = \sum_{k=1}^{D^2} A_k \rho
A_k^\dagger$, which is normally subject to a trace-preserving
condition $\sum_k A_k^\dagger A_k = 1$. We focus here on systems of
$n$ qubits so that $D=2^n$.

We start by expanding the noise operators in a basis of Pauli
operators $P_i \in \mathcal{P}_n$ consists of $n$-fold tensor
product of the usual single-qubit Pauli operators $\{ 1,X,Y,Z \}$,
giving $A_k = \sum_{i=1}^{D^2} \alpha_i^{(k)} P_i / \sqrt{D}$, where
$\alpha_i^{(k)} = \Tr[ A_k P_i]/\sqrt{D}$, and the Pauli's satisfy
the orthogonality relation $\Tr[P_i P_j] = D \delta_{ij}$.  The
Clifford group $\mathcal{C}_n$ is defined as the normalizer of the
Pauli group $\mathcal{P}_n$: it consists of all elements $U_i$ of
the unitary group $U(D)$ satisfying $ U_i P_j U_i^\dagger \in
\mathcal{P}_n $ for every $P_j \in \mathcal{P}_n$.

We can analyze the effect of the twirl $\mathcal{C}_1^{\otimes n}$
by noting that any element $C_i \in \mathcal{C}_1^{\otimes n}$ can
be expressed as $C_i = P_j Y_l $, where $P_j \in \mathcal{P}_n$ and
$Y_l \in \mathcal{Y}_1^{\otimes n}$, and where we consider as
equivalent elements of each group that differ only by a phase.
Hence, the $\mathcal{C}_1^{\otimes n}$-twirl of an arbitrary channel
$\Lambda$ consists of the action
\be
   \Lambda(\rho) \rightarrow \overline{\Lambda(\rho)} = \frac{1}{|\mathcal{C}_1^{\otimes n}|} \sum_{j=1}^{|\mathcal{S}_1^{\otimes
    n}|} \sum_{l=1}^{|\mathcal{P}_n|}
    \sum_k S_j^\dagger P_l^\dagger A_k P_l S_j \rho S_j^\dagger P_l^\dagger A_k^\dagger P_l S_j.
\ee
where $|\mathcal{C}_1^{\otimes n}| =|\mathcal{S}_1^{\otimes  n}| \;
|\mathcal{P}_n|$. The effect of the Pauli-twirl is to create the
channel $\sum_i a_i P_i \rho P_i$, where $a_i = \sum_k
|\alpha_i^{(k)}|^2/D$ are probabilities, known as a Pauli channel.
The effect of the symplectic-twirl on the Pauli channel is to map
each of the non-identity Pauli operators to a uniform sum over the 3
non-identity Pauli operators. To express this we separate out terms
according to their Pauli weight $w$, where $w \in \{0,\dots, n\} $
is the number of non-identity factors in $P_l$. We let the index
$\nu_w \in \{1, \dots, {n \choose w} \}$ count the number of
distinct ways that $w$ non-identity Pauli operators can be
distributed over the $n$ factor spaces, and the index $\mathbf{i}_w
= \{i_1, \dots, i_w \}$ with $ i_j \in \{1, 2, 3 \}$ denote which of
the non-identity Pauli operators occupies the $j$'th occupied site.
Hence we have,
\be
   \overline{\Lambda(\rho)} = \frac{1}{|\mathcal{S}_1^{\otimes n}|} \sum_{j=1}^{|\mathcal{S}_1^{\otimes
    n}|}
    \sum_{w=0}^n  \sum_{\nu_w=1}^{n\choose w} \sum_{\mathbf{i}_w=1}^{3^w}
    a_{w,\nu_w,\mathbf{i}_w} S_j^\dagger P_{w,\nu_w,\mathbf{i}_w} S_j \rho S_j^\dagger P_{w,\nu_w,\mathbf{i}_w} S_j.
\ee

Now, for any term with arbitrary but fixed $w$ and $\nu_w$, we have
the expression,
\be
  \frac{1}{|\mathcal{S}_1^{\otimes n}|}
    \sum_{\mathbf{i}_w=1}^{3^w} a_{w,\nu_w,\mathbf{i}_w}  \sum_{j=1}^{|\mathcal{S}_1^{\otimes
    n}|}
    S_j^\dagger P_{w,\nu_w,\mathbf{i}_w} S_j \rho S_j^\dagger P_{w,\nu_w,\mathbf{i}_w} S_j =
   \left( \frac{1}{3^w}  \sum_{\mathbf{i}_w}^{3^w} a_{w,\nu_w,\mathbf{i}_w} \right)  \sum_{\mathbf{j}_w=1}^{3^w}
    P_{w,\nu_w,\mathbf{j}_w} \rho P_{w,\nu_w,\mathbf{j}_w}.
\ee
Consequently we obtain,
\be \overline{\Lambda(\rho)} = \sum_{w=0}^n  \sum_{\nu_w=1}^{{n \choose w}}
    r_{w,\nu_w} \sum_{\mathbf{i}_w=1}^{3^w}
    P_{w,\nu_w,\mathbf{i}_w} \rho P_{w,\nu_w,\mathbf{i}_w}
\ee
where $
 r_{w,\nu_w} = \frac{1}{3^w}  \sum_{\mathbf{i}_w=1}^{3^w}
a_{w,\nu_w,\mathbf{i}_w}$. If we explicitly apply random qubit
permutations chosen uniformly, then the effective channel
$\overline{\Lambda}^{\Pi}$ is given by
\be \overline{\Lambda}^{\Pi}(\rho)
 = \sum_{w=0}^n  p_w
    \sum_{\nu_w=1}^{{n \choose w}}
    \sum_{\mathbf{i}_w=1}^{3^w}
    {1\over 3^w {n\choose w}}
    P_{w,\nu_w,\mathbf{i}_w} \rho P_{w,\nu_w,\mathbf{i}_w},
\ee
where
\be p_w
= 3^w\sum_{\nu_w=1}^{{n\choose w}} r_{w,\nu_w}.
\ee

This symmetrized channel can now be probed experimentally by
inputting the initial state $|0\> \equiv | 0\>^{\otimes n}$ and
performing a projective measurement in the computational basis
$|l\>$, where $l \in \{ 0, 1 \}^n $. If we distinguish outcome bit
strings only according to their Hamming weight $h \in {0, \dots,
n}$, the effect is equivalent to a random permutation of the qubits.
Observe that only Pauli $X$ and $Y$ errors will affect the Hamming
weight because Pauli $Z$ errors commute with the input state. Hence
the probability of measuring an outcome with Hamming weight $h$ is
$$u_{h} = \sum_{w=0}^n  R_{hw} p_{w}$$ where $R_{hw} = {w \choose h}
2^{h}/3^w $ gives the number of Pauli operators of weight $w$ of
which exactly $h$ are either $X$ or $Y,$ and where
\be p_w = 3^w
\sum_{\nu_w=1}^{{n \choose w}} r_{w,\nu_w}  = \sum_{\nu_w=1}^{{n
\choose w}} \sum_{\mathbf{i}_w}^{3^w} p_{w,\nu_w,\mathbf{i}_w}
\ee
is a quantity of interest, i.e., the total probability of all Pauli
errors with weight $w$. Noting that the $n \times n$  matrix
$R_{hw}$ satisfies $R_{hw} = 0$ when $h > w$ and hence is upper
triangular, estimates of the $p_{w}$ can be recovered trivially from
the measured probabilities $u_{h}$ after $n$ back-substitutions. In
another approach, if we distinguish outcome bit strings only by the
parity of a random subset of $w$ qubits, then the effect is also
equivalent to a random permutation of the qubits. Thus,
experimentally we implement $\overline{\Lambda(\rho)}$ via twirling,
but access the parameters of $\overline{\Lambda}^{\Pi}(\rho)$ by
averaging over random choices of subsets of $w$ qubits. The
probability $q_w$ that the parity of random subset of $w$ qubits is
even is related to $\langle \overline{Z^{\otimes w}}\rangle$ via
\begin{equation}
c_w \equiv \langle \overline{Z^{\otimes w}}\rangle = q_w - (1 - q_w)
= 2 q_w - 1,
\end{equation}
where we have defined the variable $c_w$ for ease of notation. In
order to analyze the information content of the $c_w$ and their
relation to the error probabilities $p_w$ it is convenient to
consider the Liouville representation of the twirled channel.

\subsection{Liouville Representation of the Twirled Channel}

Because any two operators $P_{m,\nu_m,{\bf i}_m}\in\mathcal{P}_n$
either commute or anti-commute, it follows that
\begin{equation}\label{csandps}
\overline{\Lambda}^\Pi\left(P_{m,\nu_m,{\bf i}_m}\right) = \left(
\Pr(\text{comm}) - \Pr(\text{anti-comm}) \right) P_{m,\nu_m,{\bf
i}_m} = c_m P_{m,\nu_m,{\bf i}_m},
\end{equation}
where $\Pr(\text{comm})$ ($\Pr(\text{anti-comm})$) is the
probability of the channel $\overline{\Lambda}^\Pi$ acting with a
noise operators $P_{w,\nu_w,{\bf i}_w}$ which commutes
(anti-commutes) with $P_{m,\nu_m,{\bf i}_m}$. Thus, the Pauli
operators $P_{m,\nu_m,{\bf i}_m}$ are the eigenoperators of the
channel with corresponding eigenvalues $c_m$. The eigendecomposition
of $\overline{\Lambda}^\Pi$ is given by
\begin{equation}
\overline{\Lambda}^\Pi(\rho) = \sum_{w=0}^n c_w M^c_w(\rho),
\end{equation}
where $M^c_w$ are the superoperators
\begin{equation}
M^c_w(\rho) = {1\over 2^n} \sum_{\nu_w = 0}^{{n\choose w}}\sum_{{\bf
i}_w=0}^{3^w} P_{w,\nu_w,{\bf i}_w}  \tr(P_{w,\nu_w,{\bf i}_w}
\rho).
\end{equation}
We can also rewrite the usual parameterisation of
$\overline{\Lambda}^\Pi$ as
\begin{equation}
\overline{\Lambda}^\Pi(\rho) = \sum_{w=0}^n p_w M^p_w(\rho),
\end{equation}
where $M^p_w$ are the superoperators
\begin{equation}
M^p_w(\rho) = {1\over 3^w {n \choose w}} \sum_{\nu_w = 0}^{{n\choose
w}}\sum_{{\bf i}_w=0}^{3^w} P_{w,\nu_w,{\bf i}_w} \rho
P_{w,\nu_w,{\bf i}_w}.
\end{equation}
By considering the Liouville representation of these superoperators
it is easy to show that the $M^c_w$ are orthogonal and that the
$M^p_w$ are orthogonal. Thus, $\{c_w\}_{w=0}^n$ parameterizes the
channel $\overline{\Lambda}^\Pi$  uniquely, and $\{p_w\}_{w=0}^n$
also parameterizes the same channel uniquely. Using the Liouville
representation it follows that these parameterizations are related
by a $(n+1)\times(n+1)$ matrix $\Omega$ such that
\begin{gather}
c_w = \sum_{w'=0}^n p_{w'} \Omega_{w,w'},\\
p_w = \sum_{w'=0}^n c_{w'} \Omega_{w,w'}^{-1}
\end{gather}
with $\Omega$ defined by
\begin{gather}
\Omega_{w,w'} = {4^n\over 3^{w+w'} {n\choose w}{n\choose w'}} \langle M^c_w, M^p_{w'}\rangle\label{omega1}\\
\Omega_{w,w'}^{-1} = \langle M^p_{w} , M^c_{w'}
\rangle,\label{omega1inv}
\end{gather}
where $\langle \cdot , \cdot \rangle$ is the Hilbert-Schmidt inner
product of superoperators acting on Liouville space defining the
notion of orthogonality discussed above. To obtain an explicit
expression for $\Omega_{w,w'}$, we start from \eqref{csandps} and
observe that a Pauli operator of weight $w$ is scaled by a channel
of the form
\begin{equation}
\mathcal{N}_{w'}(\rho) = {1\over3^{w'}{n\choose w'}}
\sum_{\nu_{w'}=0}^{{n\choose w'}}\sum_{{\bf i}_{w'}=0}^{3^{w'}}
P_{w',\nu_{w'},{\bf i}_{w'}} \rho P_{w',\nu_{w'},{\bf i}_{w'}}.
\end{equation}
This implies
\begin{equation}
\Omega_{m,w} = -1 + \sum_{L=\max(0, {w + m - n})}^{\min(m,w)} {{n -
m\choose w - L} {m\choose L} \over {n\choose w}} {3^L + (-1)^L\over
3^L},
\end{equation}
and, using \eqref{omega1} and \eqref{omega1inv}, it follows that
\begin{equation}\label{omegainv}
\Omega_{m,w}^{-1} = { 3^{m + w} {n\choose m}{n\choose w} \over 4^{n}
} \Omega_{m,w}.
\end{equation}

\subsection{Simple examples}

For the case of two-qubit channels, this matrix is given by
\begin{gather}
\Omega = \left(
\begin{array}{rrr}
1 & 1 & 1\\
1 & {1\over 3}  & -{1\over 3}\\
1 & -{1\over 3} & {1\over 9}
\end{array}
\right)\\
\Omega^{-1} = {1\over 16} \left(
\begin{array}{rrr}
1 &   6 &   9\\
6 &  12 & -18\\
9 & -18 &   9
\end{array}
\right),
\end{gather}
and for the case of three-qubit channels, it is given by
\begin{gather}
\Omega = \left(
\begin{array}{rrrr}
1 & 1 & 1 & 1\\
1 &  {5\over 9} &  {1\over 9}  & -{1\over 3}\\
1 &  {1\over 9} & -{5\over 27} &  {1\over 9}\\
1 & -{1\over 3} &  {1\over 9}  & -{1\over 27}
\end{array}
\right)\\
\Omega^{-1} = {1\over 64} \left(
\begin{array}{rrrr}
1  &   9  &  27  &  27\\
9  &  45 &   27  & -81\\
27 &  27 & -135  &  81\\
27 & -81 &   81  & -27
\end{array}
\right)
\end{gather}

\section{Appendix B: Uncorrelated Noise Locations}

A noise channel over $n$ qubits that has a distribution of error
locations which is uncorrelated, but otherwise arbitrary, is mapped
under twirling and random permutations to a channel which is a
tensor product of $n$ single-qubit depolarizing channels. Each of
these single-qubit channels has the form
\begin{equation}
{\mathcal D}(\rho) = (1-p)\rho + {p\over 3}(X\rho X+Y\rho Y+ Z\rho
Z)
\end{equation}
and scales a single-qubit Pauli operator by $c_1=1-{4\over3}p$.
Thus, the $n$ qubit channel will scale a Pauli operator with weight
$w$ by $c_w=c_1^w$.

Due to the finite accuracy with which the eigenvalues $c_w$ are
estimated through experiment, we can only impose this as a necessary
condition for the independence of the distribution of error
locations. Therefore, any estimate of the eigenvalues which makes
such an exponential dependence unlikely, also implies that the
distribution of the error locations is unlikely to be uncorrelated.

\section{Appendix C: Statistical Analysis}

The circuit complexity is depth $2$ with only $2n$ single-qubit
gates required for the protocol. The outcome from any single
experiment is just a binary string.  The number of such trials
required to estimate the probability $q_{w'}$ of even parity for a
random subset of $w'$ bits to within a given precision $\delta$ is
clearly independent of the number of qubits because the problem is
reduced to the simple task of estimating the probability of a
2-outcome classical statistical test. More precisely from the
Chernoff inequality, any estimate of the exact average
$\mathbb{E}[X] = q_w$ after $K$ \emph{independent trials} satisfies,
\begin{equation}
\mathrm{Pr}(| \frac{1}{K} \sum_{i=1}^K X_i - \mathbb{E}[X]| > \delta
) \leq 2\exp(-\delta^2 K).
\end{equation}
We see that the number of experiments required to estimate $q_w$ to
precision $\delta$ with constant probability is at most,
$$ K  = \log(2)\delta^{-2},$$
where each experiment is an \emph{independent trial} consisting of a
single shot experiment in which the Clifford gates are chosen
uniformly at random. The number of experimental trials required to
estimate the complete set of probabilities $\{ q_0, \dots,q_w, \dots
q_n \}$  can be obtained from the union bound,
\begin{equation}
\mathrm{Pr}( \cup_w \E_w) \leq \sum_w \mathrm{Pr}(\E_w)
\end{equation}
which applies for arbitrary events $\E_w$. In our case each $\E_w$
is associated with the event that $| \frac{1}{K} \sum_{k=1}^K X_k -
\mathbb{E}[X]| > \delta$ and similarly $\cup_w \E_w$ is the
probability that at least one of the $n+1$ estimated probabilities
$q_w$ satisfies this property (i.e., is an unacceptable estimate)
after $K$ trials. Whence, the probability that at least one of the
$n+1$ estimated probabilities is outside precision $\delta$ of the
exact probability is bounded above by,
$$\mathrm{Pr}( \cup_w \E_w)  \leq 2(n+1) \exp(-\delta^2 K). $$
This implies that at most $K = \mathcal{O}(\delta^{-2}\log(2(n+1)))$
experimental trials are required to estimate each of the components
of the (probability) vector $(q_0, \dots, q_n)$ to within precision
$\delta$ with constant probability.

\subsection{Uncertainty of $p_w$ Estimates}

Given an estimate of the $c_w=\langle Z^{\otimes w}\rangle$ with
some variance $\sigma^2$, the variance of the estimate of a
particular $p_w$ is given by
\begin{equation}
\sigma_w^2 = \sum_{i=0}^{n} \sum_{j=0}^n
\Omega_{w,i}^{-1}\Omega_{w,j}^{-1}\Cov(c_i,c_j)
\end{equation}
Assuming the estimates for all $c_w$ have the same variance, the
positivity constraint on the covariance matrix of the $c_w$
estimates requires that $|\Cov(c_i,c_j)|\le \sigma^2$, yielding the
upper bound
\begin{equation}
\sigma_w^2 \le \sigma^2 \sum_i \sum_j
\left|\Omega_{w,i}^{-1}\Omega_{w,j}^{-1}\right|.
\end{equation}
From \eqref{csandps}, it is clear that
\begin{equation}
|\Omega_{w,w'}|\le 1,
\end{equation}
so the from \eqref{omegainv} we have
\begin{equation}
\sigma_w \le \sigma 3^{w} {n \choose w}.
\end{equation}
Using the fact that
\begin{equation}
{n\choose w} \le {n^w \e^w \over w^w},
\end{equation}
we can rigorously show that the uncertainty $\sigma_w$ on the
estimate of $p_w$ is bounded by
\begin{equation}
\sigma_w \le \sigma \left({3\over w}\right)^{w} n^{w} \e^w.
\end{equation}

\begin{figure}
\includegraphics[width=7in]{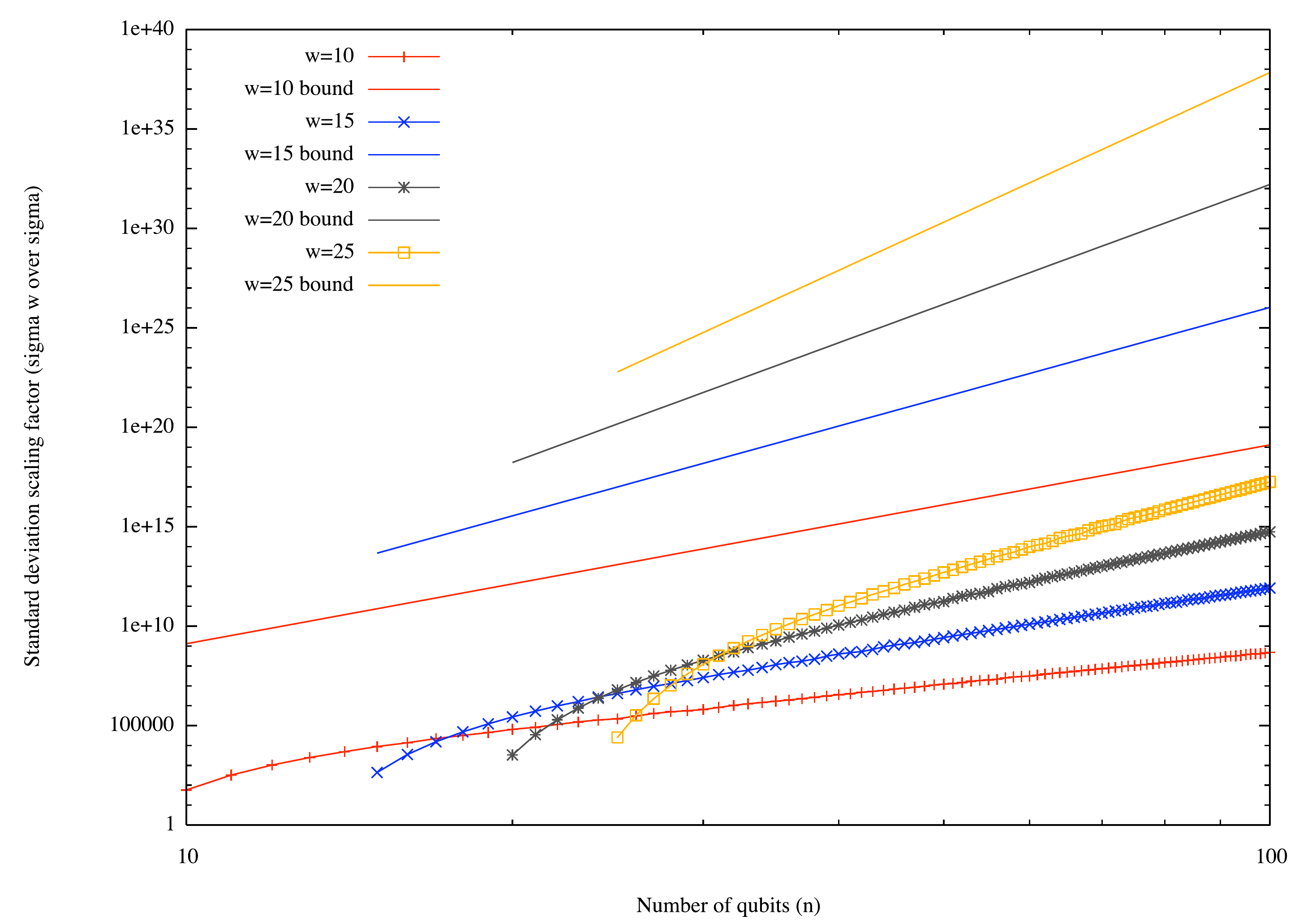}
\caption{Analytic upper bound on standard deviation scaling factor,
and numerical calculations for scaling factor with worst case
correlations.\label{fig:upper-bound}}
\end{figure}

Exact numerical computation of the uncertainty scaling factor
${\sigma_w\over\sigma}$, depicted in Figure~\ref{fig:upper-bound},
indicates that for fixed $w$ the uncertainty grows as a polynomial
in $n$, but the degree of that polynomial depends linearly on $w$.
For large $n$, we find that
\begin{equation}
{\sigma_{w}\over\sigma} \approx \e^{k w + a} n^{m w + b},
\end{equation}
where
\begin{align}
k & = -1.62521  \pm 0.09716\\
a & = 11.854    \pm 1.785 \\
m & = 0.638924  \pm 0.0089\\
b & = -0.982147 \pm 0.1647.
\end{align}

\section{Appendix D: Methods}

The two-qubit liquid-state experiments were performed on a sample
made from 10mg of $^{13}$C labeled chloroform  (Cambridge Isotopes)
dissolved in 0.51ml of deuterated acetone.  The experiment was
performed on a 700MHz Bruker Avance spectrometer using a dual
inverse cryoprobe.  The pulse programs were optimized on a
home-built pulse sequence compiler which pre-simulates the pulses in
an efficient pairwise manner and takes into account first order
phase and coupling errors during a pulse by modifications of the
refocussing scheme and pulse phases \cite{benchmark}.
The solid-state experiments were performed on a single crystal of
malonic acid which contained $\approx $7\% triply labeled $^{13}$C
molecules \cite{Baugh06a}. The experiments were performed at room
temperature with a home-built probe. Apart from an initial
polarization transfer, the protons were decoupled using the SPINAL64
sequence \cite{SPINAL64}. The required control fields that
implemented the unitary propagators and state-to-state
transformations were found using the GRAPE optimal control method
\cite{GRAPE} and made robust to inhomogeneities in both the r.f. and
static fields.  The implemented versions of the pulses were
corrected for non-linearities in the signal generation and
amplification process through a pickup coil to measure the r.f.
field at the sample and a simple feedback loop. The error
probabilities for each experiment were calculated using a
constrained maximum likelihood function.

\textbf{Acknowledgements} This work benefitted from discussions with
R. Blume-Kohout, R. Cleve, D. Gottesman, E. Knill, B. Levi, and A.
Nayak and technical expertise from M. Ditty. This research was
supported by NSERC, MITACS, ORDCF, ARO and DTO.





\begin{thebibliography}{99}


\bibitem{Sho95a} Shor, P.~W. Scheme for reducing decoherence in quantum computer memory. \emph{Phys. Rev. A} \textbf{52}, R2493 (1995).

\bibitem{Ste96a} Steane, A.~M. Error correcting codes in quantum theory. \emph{Phys. Rev. Lett}. \textbf{77}, 793 (1996).

\bibitem{shor:qc1996a}
Shor, P.~W., Fault-tolerant quantum computation. {\em Proceedings of
the Symposium on the Foundations of Computer Science, } 56-65 (IEEE
press, Los Alamitos, California, 1996).

\bibitem{aharonov:qc1996a}
Aharonov, D. and Ben-Or, M., Fault-tolerant quantum computation with
constant error. {\em Proceedings of the 29'th Annual ACM Symposium
on the Theory  of Computing, }  176-188 (ACM Press, New York, New
York, 1996).

\bibitem{kitaev:qc1997a}
Kitaev, A.~Y. Quantum computations: algorithms and error correction.
{\em Uspekhi Mat. Nauk}{ \bf 52}, 53-112 (1997).

\bibitem{knill:qc1998a}
Knill, E., Laflamme, R., and Zurek, W.~H. Resilient quantum
computation. {\em Science} {\bf 279}, 342-345 (1998).

\bibitem{NC-QPT} Chuang, I. and Nielsen, M. \emph{J. Mod. Opt}. \textbf{44}, 2455 (1997).

\bibitem{AAPT} D'Ariano, G. M. and Lo Presti, P. \emph{Phys. Rev. Lett}. \textbf{86}, 4195
(2001).

\bibitem{DC} Mohseni, M. and Lidar, D. \emph{Phys. Rev. Lett}. \textbf{97}, 170501 (2006).

\bibitem{8qubit} Haffner, H. \emph{et al}. \emph{Nature} \textbf{438}, 643
(2005).

\bibitem{Mqubits} Leibfried, D. \emph{et al}. \emph{Nature} \textbf{438}, 639 (2005).

\bibitem{12qubit} Negrevergne, C. \emph{et al}.
 Benchmarking quantum control methods on a 12-qubit
 system.
 \emph{Phys. Rev. Lett.} \textbf{96}, 170501 (2006).

\bibitem{Emerson03}
Emerson, J. \emph{et al}. Pseudo-Random Unitary Operators for
Quantum Information Processing.
\emph{Science} \textbf{302}: 2098-2100 (2003).

\bibitem{LopezLevi06} Levi, B. \emph{et al}.
Efficient Error Characterization in Quantum Information Processing,
 \emph{Phys. Rev. A} \textbf{75}, 022314 (2007).

\bibitem{EAZ05}
Emerson, J., Alicki,  R.,  Zyczkowski, K. \emph{J. Opt. B: Quantum
and Semiclassical Optics}, {\bf 7} S347-S352 (2005).

\bibitem{DCEL} Dankert, C. \emph{et al}.
submitted to \emph{Phys. Rev. Lett}., quant-ph/0606161 (2006).

\bibitem{Cirac05} Dur, W. \emph{et al}. 
\emph{Phys. Rev. A} \textbf{72}, 052326 (2005).

\bibitem{NMRQIP} Cory, D.G. \emph{et al}.
NMR Based Quantum Information Processing: Achievements and
Prospects. \emph{Fortschritte der Physik}  \textbf{48}, 875 - 907
(2000).

\bibitem{BDSW} Bennett, C. \emph{et al}.
\emph{Phys. Rev. A} \textbf{54}(5), 3824-3851 (1996).

\bibitem{AGP} Aliferis, P., Gottesman, D., Preskill, J. Quantum accuracy threshold for concatenated distance-3
codes. \emph{Quant. Inf. Comp}. \textbf{6}, 97-165 (2006

\bibitem{Fortunato02} Fortunato, E. M. \emph{et al}.
J. Chem. Phys. 116, 7599-7606 (2002)

\bibitem{NielsenFidelity} Nielson, M., \emph{Phys. Lett. A} \textbf{303} (4): 249-252
(2002).

\bibitem{knill-nature}  Knill, E. Scalable quantum computing in the presence of large
detected-error rates. \emph{Nature} \textbf{434}, 39-44 (2005).

\bibitem{knill-PRA}
Knill, E. Quantum computing with realistically noisy devices.
\emph{Phys. Rev. A} \textbf{71}, 042322 (2005).

\bibitem{Knill} The latter approach was suggested to us by E. Knill.






\bibitem{C48} Cory, D.G., Miller, J.B., and Garroway, A.N.,
Time-Suspension Multiple- Pulse Sequences: Applications to
Solid-State Imaging. \emph{Journal of Magnetic Resonance}
\textbf{90}, 205-213 (1990).

\bibitem{WaHa} Haeberlen, U. \emph{Advances in Magnetic Resonance}, Ed. J. Waugh, Academic Press, New York (1976).


\bibitem{Boulant} Boulant, N. \emph{et al.} \emph{Phys. Rev. A} \textbf{67}, 042322 (2003).







\bibitem{benchmark} Knill, E. \emph{et al}.
\emph{Nature} 404, 368-370 (2000).

\bibitem{Baugh06a} Baugh, J. \emph{et al}.
\emph{Phys. Rev. A} \textbf{73}, 022305 (2006).

\bibitem{SPINAL64} Fung, B.M., Khitrin, A.K., Ermolaev, K., \emph{Journal of
Magnetic Resonance} \textbf{142}, 97-101 (2000).

\bibitem{GRAPE} Khaneja, N., Reiss, T., Kehlet, C., Herbruggen,
T.S., Glaser, S.J. \emph{Journal of Magnetic Resonance}
\textbf{172}, 296-305 (2005).


\end{thebibliography}
\end{document}